\documentclass[12pt,a4paper]{article}
\usepackage[T1]{fontenc}
\usepackage{physics}
\usepackage[utf8]{inputenc}
\usepackage[affil-it]{authblk} 
\usepackage{etoolbox}
\usepackage{lmodern}
\usepackage{xcolor}
\usepackage{graphicx}
\usepackage{titletoc}

\usepackage{xr}
\usepackage{lineno}

\usepackage{float}
\usepackage[font=small,labelfont=bf]{caption}
\usepackage{hyperref}
\hypersetup{
    colorlinks=true,
    linkcolor=blue,
    citecolor=blue,
    filecolor=blue,      
    urlcolor=blue,
    pdftitle={Overleaf Example},
    pdfpagemode=FullScreen,
    }

\usepackage{geometry}
\usepackage{pdfpages} 
\usepackage{pgffor} 

\makeatletter
\AtBeginDocument{\let\LS@rot\@undefined}
\makeatother

\def\supplementfilename{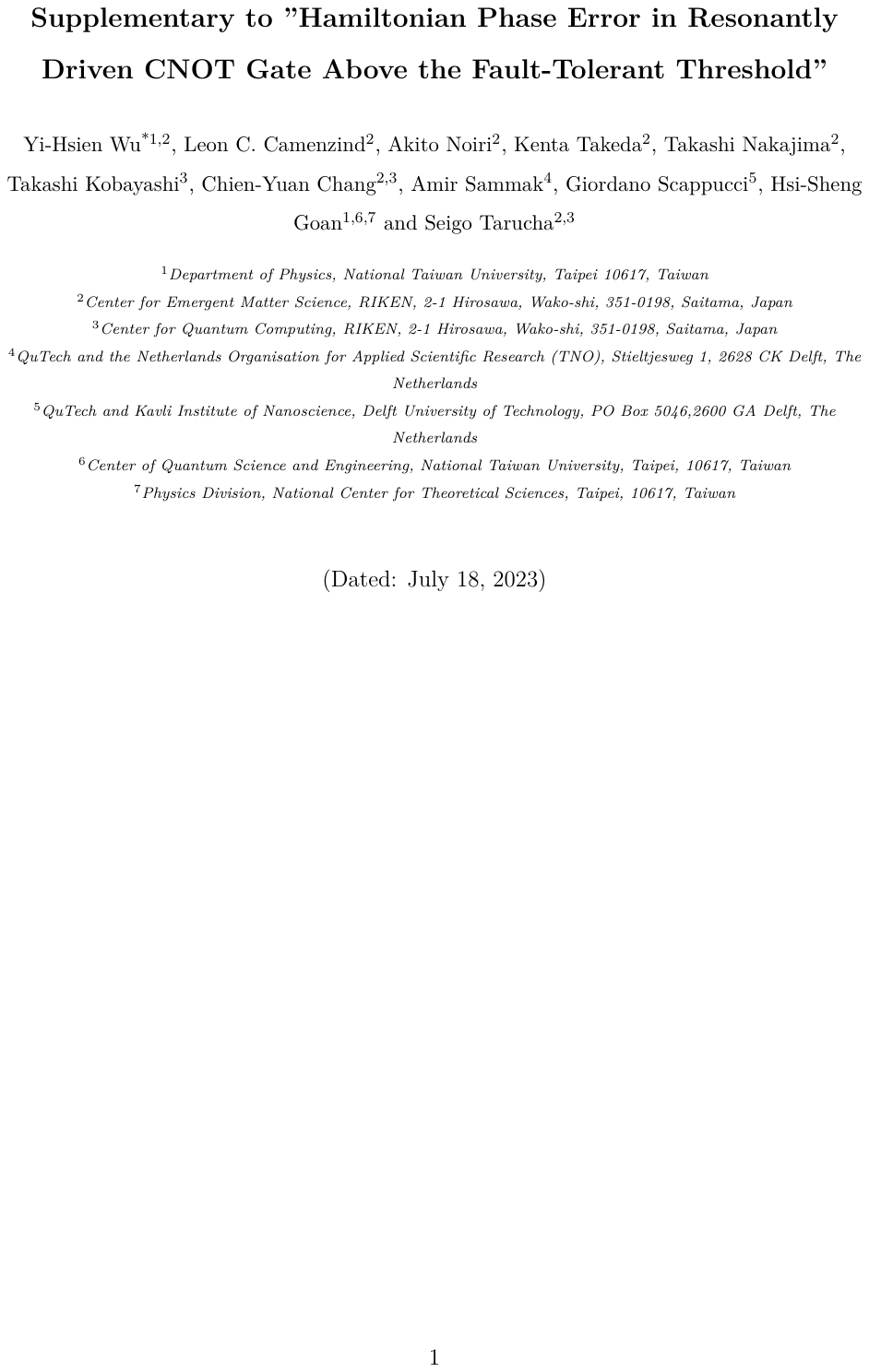}

\pdfximage{\supplementfilename}
\def\numbersupplementpages{\the\pdflastximagepages}

\newif\ifarXiv
\arXivtrue 

\title{\large\textbf{Hamiltonian Phase Error in Resonantly Driven CNOT Gate Above the Fault-Tolerant Threshold} }
\author[*1,2]{Yi-Hsien Wu}
\author[2]{Leon C. Camenzind}
\author[2]{Akito Noiri}
\author[2]{Kenta Takeda}
\author[2]{Takashi Nakajima}
\author[3]{Takashi Kobayashi}
\author[2,3]{Chien-Yuan Chang}
\author[4]{Amir Sammak}
\author[5]{Giordano Scappucci}
\author[1,6,7]{Hsi-Sheng Goan}
\author[2,3]{Seigo Tarucha}

\affil[1]{Department of Physics, National Taiwan University, Taipei 10617, Taiwan}
\affil[2]{Center for Emergent Matter Science, RIKEN, 2-1 Hirosawa, Wako-shi, 351-0198, Saitama, Japan}

\affil[3]{Center for Quantum Computing, RIKEN, 2-1 Hirosawa, Wako-shi, 351-0198, Saitama, Japan}

\affil[4]{QuTech and the Netherlands Organisation for Applied Scientific Research (TNO), Stieltjesweg 1, 2628 CK Delft, The Netherlands}

\affil[5]{QuTech and Kavli Institute of Nanoscience, Delft University of Technology, PO Box 5046,2600 GA Delft, The Netherlands}

\affil[6]{Center of Quantum Science and Engineering, National Taiwan University, Taipei, 10617, Taiwan}
\affil[7]{Physics Division, National Center for Theoretical Sciences, Taipei, 10617, Taiwan}

\date{\normalsize (Dated: \today)}

\newcommand{\rev}[1]{{\color{black}#1}}   
\newcommand{\lcc}[1]{{\color{black}#1}}   

\begin{document}
\maketitle

\begin{abstract}
\small
Because of their long coherence time and compatibility with industrial foundry processes, electron spin qubits are a promising platform for scalable quantum processors. A full-fledged quantum computer will need quantum error correction, which requires high-fidelity quantum gates. \rev{Analyzing and mitigating the gate errors are useful to improve the gate fidelity.} Here, we demonstrate a simple yet reliable calibration procedure for a high-fidelity controlled-rotation gate in an exchange-always-on Silicon quantum processor allowing operation above the fault-tolerance threshold of quantum error correction. We find that the fidelity of our uncalibrated controlled-rotation gate is limited by coherent errors in the form of controlled-phases and present a method to measure and correct these phase errors. We then verify the improvement in our gate fidelities by randomized benchmark and gate-set tomography protocols. Finally, we use our phase correction protocol to implement a virtual, high-fidelity controlled-phase gate.
\end{abstract}

\section{Introduction}\label{introduction}
Spin qubits in solid state devices \cite{loss_quantum_1998} are a promising platform for large-scale quantum computers. Universal control has recently been demonstrated in a six qubit device in Silicon \cite{philips_universal_2022}, and a four qubit device in Germanium \cite{hendrickx_four-qubit_2021}, marking a first step of scaling up spin qubit devices. Spin qubits in Silicon exhibit long coherence times \cite{yoneda_quantum-dot_2018, veldhorst_addressable_2014}, fast manipulation \cite{yoneda_fast_2014} and ability to operate at an elevated temperature \cite{yang_operation_2020, petit_universal_2020,camenzind_spin_2022}. The compatibility with the already matured semiconductor industry processes allows mass fabrication of devices \cite{zwerver_qubits_2022}, integration with cryo-electronics \cite{xue_cmos-based_2021}, and opens the potential for high-performance integrated quantum circuits in the future \cite{Vandersypen_interfacing_2017}. Quantum error correction, a critical feature of large-scale quantum computers, has also been demonstrated in a \rev{three-qubit} device recently \cite{takeda_quantum_2022}. These \lcc{progresses} make spin qubits a viable qubit platform for the future.

\rev{To implement} large-scale quantum computers the ability to implement quantum error correction code \rev{is required}. One of the most promising quantum error correction codes is the surface code \cite{fowler_surface_2012}. Typically, under certain assumptions of the error model, the surface code gives an error threshold of 1\% \cite{wang_surface_2011}. High fidelity single-qubit \cite{yoneda_quantum-dot_2018, yang_silicon_2019} and two-qubit gates \cite{noiri_fast_2022, xue_quantum_2022, mills_two-qubit_2022} which satisfy this error threshold have been demonstrated with spin qubits in isotopically enriched Silicon. Among these results, the two-qubit gates are implemented as a controlled-phase (CZ) gate or controlled-rotation (CROT) gate. A high-fidelity CZ gate requires fast and precise pulses to control the exchange coupling between two qubits. The CROT gate, on the other hand, can be implemented in a less demanding way by keeping the always-on exchange \cite{noiri_fast_2022,huang_fidelity_2019}. In the exchange-always-on system, the CROT gate fidelity is reduced by a coherent off-resonant Hamiltonian phase error which has the form of a controlled-phase. This phase error must be mitigated to obtain high-fidelity CROT gates above the fault-tolerant threshold. Previous work avoids this problem by shifting control microwave frequecy \cite{noiri_fast_2022}. The microwave frequency is adjusted by a feed-back loop \rev{to minimize this phase error}. Here we systematically compensate the effect of these phase errors by shifting the phase of the applied microwave pulses. We measure the phase errors with a calibration sequence and compensate its effects. Our procedure to compensate these controlled-phase errors enables us to implement a CZ gate virtually, similar to a virtual single-qubit z-gate \cite{mckay_efficient_2017}, thus without additional execution time in the quantum circuit. The ability to implement both high-fidelity CROT and a virtual CZ gate without complicated pulse engineering makes the exchange-always-on system interesting to study. Compared to a synthesized implementation using CZ gates \cite{xue_quantum_2022, mills_two-qubit_2022}, the CROT gate allows for a native, resonant CNOT logical gate with a fidelity above the fault tolerant threshold \cite{noiri_fast_2022}, which makes this gate relevant for future spin based quantum processors.

Here, we demonstrate a procedure to obtain a high-fidelity resonantly driven CROT gate in an exchange-always-on two-qubit system. We first present a systematic way to measure the accumulated phase error and then a method to compensate for these gate errors. We use randomized benchmarking (RB) protocol \cite{magesan_scalable_2011, magesan_efficient_2012} to compare the gate fidelity with and without compensation. We then perform gate-set tomography (GST) \cite{nielsen_gate_2021} to obtain the details on the error processes of our quantum gates using experimental and simulated data. The experimental and simulation data results show good agreement, which \rev{proves the validity of }the quantum gate model we use for simulation. Finally, we demonstrate the implementation of a virtual high-fidelity CZ gate using the compensation method and benchmark the performance of this virtual CZ gate with GST.

\section{Results}\label{results}

\subsection{Device and controlled rotation gates}\label{result:device}
\begin{figure}[H]
\centering
\captionsetup{font=footnotesize,skip=0pt,width=1\linewidth}
\includegraphics[width=1\textwidth]{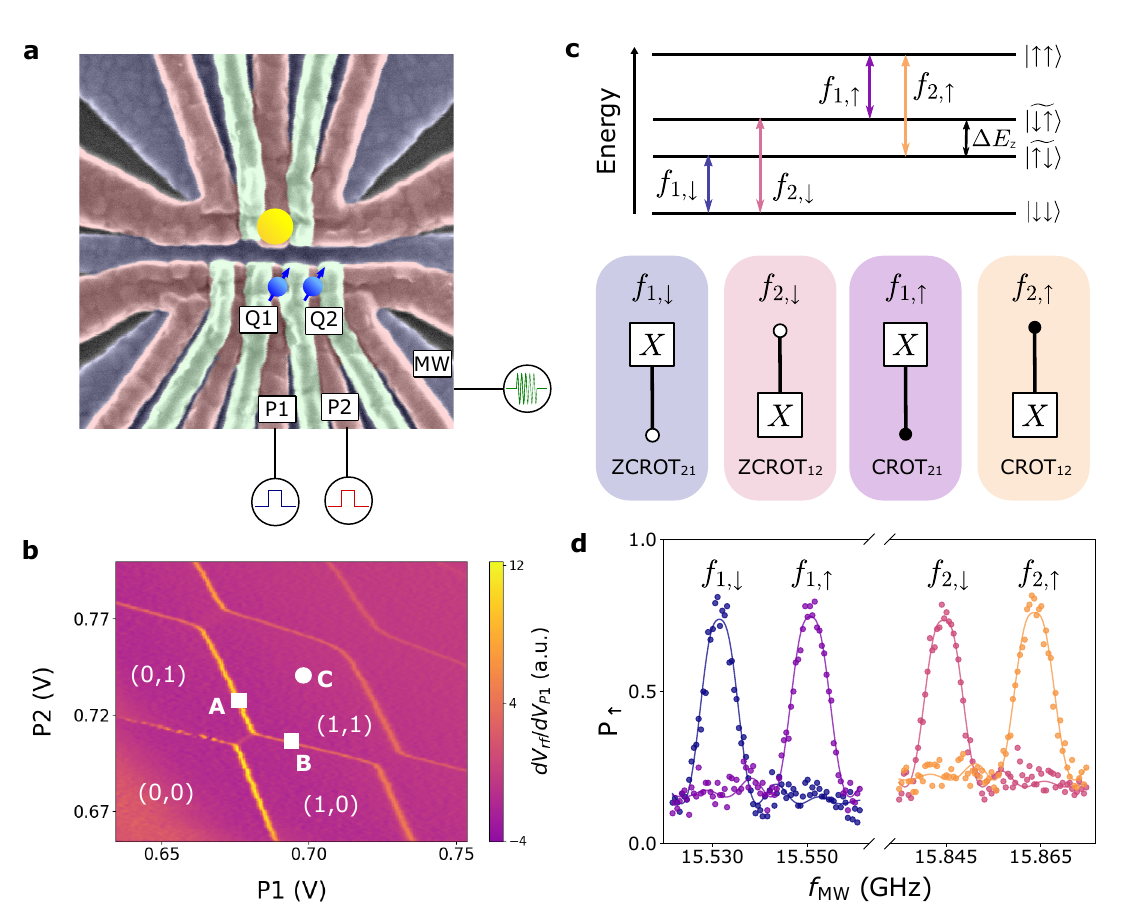}
\caption{\textbf{Two qubit system.} 
\textbf{a} False-colour scanning electron microscope image of a device identical to the one measured. The two quantum dots are formed below plunger gate electrodes P1 ($Q_{1}$) and P2 ($Q_{2}$) and the yellow circle indicates the charge sensor quantum dot. Quantum gates are implemented via electric dipole spin resonance in the gradient field of a micro-magnet (not shown) by applying microwave signals to the MW gate electrodes. 
\textbf{b} Charge stability diagram around the qubit operation condition. The number of electrons in two quantum dots is denoted as $(N_{1},N_{2})$. Readout and initialization of qubit $Q_{1}$ ($Q_{2}$) is performed at square A (B). Qubit operations are executed at the charge symmetry point (circle C) to achieve high-fidelity two-qubit gates.
\textbf{c} Energy level diagram of the two-qubit system. Exciting one of the four frequencies rotates the target qubit conditioned on the state of the other qubit, which allows the implementation of controlled rotation (CROT) and zero-controlled rotation (ZCROT) gates. The label $X$ indicates a $\pi/2$ pulse around the x-axis of the qubit Bloch-sphere.
\textbf{d} Electric dipole spin resonance peaks. The measured spectra shows the transition frequencies of $Q_{1}$ when $Q_{2}$ is in $\ket{\downarrow}$ (blue) and $\ket{\uparrow}$ (purple) and frequencies of $Q_{2}$ when $Q_{1}$ is in $\ket{\downarrow}$ (red) and $\ket{\uparrow}$ (orange). }\label{fig1}
\end{figure}

Fig.~\ref{fig1} (a) shows the device used for the experiment, which is a triple quantum dot device fabricated in isotopically purified silicon quantum well, the same device as used in Ref.~\cite{noiri_fast_2022}. A three-layer aluminum gate stack is deposited to fabricate the gate electrodes, which control the electric confinement potential of the quantum dots. A cobalt micro-magnet is deposited on the gate stack to achieve a gradient magnetic field. The gradient magnetic field allows individual addressing of the spins in the quantum dots and manipulating the spin qubit state by performing electric dipole spin resonance (EDSR) with gradient magnetic field generated by the micro-magnet. The device has a charge sensor quantum dot in the upper part of the device and an array of three quantum dots in the lower part. We perform charge sensing with reflectometry \cite{noiri_radio-frequency-detected_2020} and accumulate an electron in the center (qubit $Q_{1}$) and right dot (qubit $Q_{2}$), while the leftmost dot is used as an extension of the left reservoir. Fig.~\ref{fig1} (b) shows the stability diagram around this configuration. Energy selective single-shot readout is used for qubit readout and initialization \cite{elzerman_single-shot_2004}. 

As the exchange coupling between the two qubits is turned on, when \rev{the Zeeman energy difference $\delta E_{\text{z}}$ between two qubits is much larger than the exchange coupling $J$}, the energy levels of $\ket{\uparrow\downarrow}$ and $\ket{\downarrow\uparrow}$ are lowered, and the basis states that diagonalize the Hamiltonian become $\ket{\uparrow\uparrow}$, $\widetilde{\ket{\uparrow\downarrow}}$  $\widetilde{\ket{\downarrow\uparrow}}$ and $\ket{\downarrow\downarrow}$. The two-qubit Hamiltonian diagonalized by these states is \cite{huang_fidelity_2019, russ_high-fidelity_2018}
\begin{equation}\label{eq:H_lab}
H(t) = \frac{h}{2}\begin{pmatrix}
2\bar{E_{\text{z}}} & B(t) & B(t) & 0 \\ 
B^{*}(t) & \delta \tilde{E_{\text{z}}} -J & 0 & B(t) \\
B^{*}(t) & 0 & -\delta \tilde{E_{\text{z}}} - J & B(t) \\
0 & B^{*}(t) & B^{*}(t) & -2\bar{E_{\text{z}}}
\end{pmatrix},
\end{equation}
where $\bar{E_{\text{z}}}$ is the averaged Zeeman energy of the two qubits, \rev{$\delta\tilde{E}_{\text{z}}=\sqrt{J^{2} + \delta E_{\text{z}}^{2}}$ the effective Zeeman energy difference and $B(t)$ the effective magnetic field induced by the EDSR.} Fig.~\ref{fig1} (c) shows the energy levels of the four basis states. \rev{This results in four distinct transition resonance frequencies $f_{m,\sigma}=\bar{E_{\text{z}}} + (c_{\sigma} J + c_{m} \delta\tilde{E}_{\text{z}})/2$ where $m=1,2$ is the qubit index and $\sigma=$ $\downarrow,\uparrow$ is the spin with the coefficients $c_{1}=c_{\downarrow}=-1, c_{2}=c_{\uparrow}=+1$}. By exciting one of the four transition frequencies, we implement the resonantly driven zero-controlled rotations (ZCROT) and controlled rotations (CROT) \cite{zajac_resonantly_2018}. The ZCROT rotates the target qubit if the control qubit is in $\ket{\downarrow}$ (0 state), and the CROT rotates the target qubit if the control qubit is in $\ket{\uparrow}$ (1 state). The notation CROT$_{\text{ctrl, targ}}$ (ZCROT$_{\text{ctrl, targ}}$) indicates a rotation of the target qubit \rev{($\text{targ}=1,2$)} if the control qubit \rev{($\text{ctrl}=1,2$)} is in $\ket{\uparrow}$ ($\ket{\downarrow}$) state. Fig.~\ref{fig1} (d) shows the measured EDSR frequencies and $\delta  E_{\text{z}}\sim 310$ MHz. We choose $J = 18$ MHz such that the system is in an optimal condition for high two-qubit gate fidelities \cite{noiri_fast_2022}.

The effective magnetic field induced by the EDSR has the form $B(t)=f_{R} e^{2i\pi f_{\text{MW}}t}$ with $f_{\text{MW}}$ the microwave driving frequency and $f_{R}$ the Rabi frequency. To implement the CROT$_{12}$ gate, where $Q_{1}$ is the control qubit, and $Q_{2}$ is the target qubit, we choose a driving frequency that is resonant with the corresponding transition frequency, i.e., $f_{\text{MW}}=f_{2,\uparrow}$. By substituting $B(t)$ into the Hamiltonian given in Eq.~(\ref{eq:H_lab}), transforming to the rotating frame (see Methods~\ref{methods:sim}) and neglecting far off-resonance terms using the rotating wave approximation (RWA), the Hamiltonian becomes \cite{huang_fidelity_2019,russ_high-fidelity_2018}

\begin{align}\label{eq:H_rot}
H_{\text{CROT}_{12}}(t) =  \frac{h}{2}\begin{pmatrix}
0 & f_{R}  & 0 & 0 \\
f_{R}  & 0 & 0 & 0 \\
 0 & 0 & 0 & f_{R} e^{-2i\pi Jt} \\
 0 & 0 & f_{R} e^{2i\pi Jt} & 0 
 \end{pmatrix}.
\end{align}
The upper-left 2-by-2 sub-block provides the desired controlled rotation, while the lower-right 2-by-2 sub-block introduces error to the \lcc{gate}. Choosing $f_{R}=J/\sqrt{15}$ cancels out the population transfer caused by the lower-right sub-block \rev{for the $\pi$ and the half-$\pi$ CROT}, but two $z$-phases resulting from $e^{\pm 2i\pi Jt}$ terms will be accumulated in this sub-block. This results in a controlled-phase error which accumulates in the $\ket{\downarrow\downarrow}$ and $\widetilde{\ket{{\downarrow\uparrow}}}$ states \cite{huang_fidelity_2019,russ_high-fidelity_2018}. We verify this source of error using GST experiments as we discuss later. We call these phase errors the off-resonant Hamiltonian phase errors since they are errors arising from the control Hamiltonian. \rev{We change the rotating frame by offsetting the microwave phase in the pulse sequence to account for the accumulated phase errors. This allows us to correct these off-resonant Hamiltonian phase errors.}

\subsection{Measuring the off-resonant Hamiltonian phase error}\label{result:measuring}

\begin{figure}[H]
\centering
\captionsetup{font=footnotesize,skip=0pt,width=1\linewidth}
\includegraphics[width=1\textwidth]{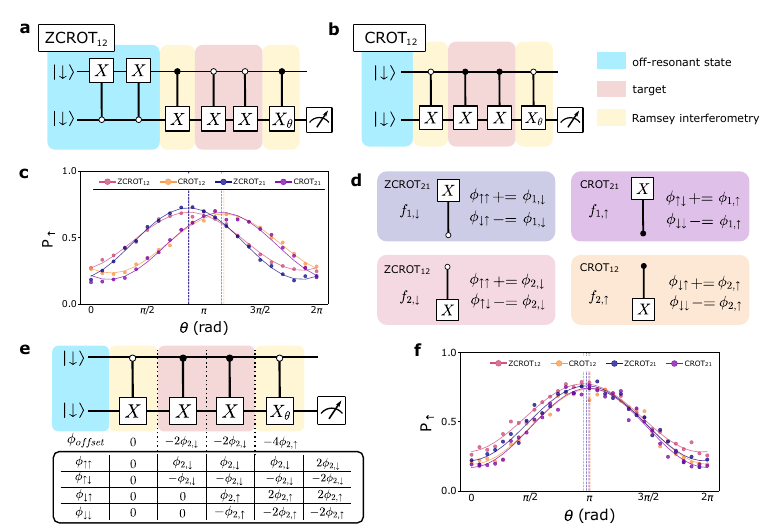}
\caption{\textbf{Calibration sequences for measuring the off-resonant Hamiltonian phase error.} 
\textbf{a, b} Calibration sequences used to measure off-resonant Hamiltonian phase errors for ZCROT$_{12}$ and CROT$_{12}$, where X denotes a half-$\pi$ rotation. For all the sequences see Supplementary Fig.~S1.
\textbf{c} Phase shifts in the calibration sequences. Spin up probability measured in each calibration sequences are fitted with $A\cos(\theta+\theta_{i})+B$ to obtain the phase shifts. With $A$ the amplitude, $B$ the offset, $\theta$ the phase of the second half-$\pi$ pulses and $\theta_{i}$ the phase shift. 
\textbf{d} The off-resonant Hamiltonian phase error associated with each pulse. The phase error $\phi_{m,\sigma}$ associated with each transition frequency $f_{m,\sigma}$ will accumulate in the off-resonant state when applying each pulse. \rev{Here we introduce the ${\mathbin{{+}{=}}}$ ($\mathbin{{-}{=}}$) operators which add (subtract) the value on the right to (from) the variable on the left, a common syntax in modern programming languages.}
\textbf{e} Schematic of compensation procedure for the off-resonant Hamiltonian phase error for the example of the calibration sequence for CROT$_{12}$. 
The phases accumulated on each basis states before the pulse is applied are shown in the columns. We then use this table to calculate the phase offset $\phi_{offset}$ and subtract this offset from the applied pulses' microwave phase to compensate for the effect of off-resonant Hamiltonian phase error.
\textbf{f} The measured phase shifts in the calibration sequences after the phase compensation\lcc{, demonstrating a significant improvement to the uncompensated case shown in (b).}
}\label{fig2}
\end{figure}

\rev{There are four calibration sequences used to measure the off-resonant Hamiltonian phase errors, one for each controlled rotation pulse (see Methods~\ref{methods:calibration_seq}).} \rev{Two of the calibration sequences for control qubit $Q_{1}$ and target qubit $Q_{2}$ are shown in Fig.~\ref{fig2} (a) and (b). For target qubit $Q_{1}$, the roles of $Q_{1}$ and $Q_{2}$ are swapped}. The calibration sequence has three parts: First, the control qubit is prepared to the off-resonant state of the target gate, $\ket{\uparrow}$ for the ZCROT and $\ket{\downarrow}$ for the CROT. Next, we perform a Ramsey experiment with the target sequence inserted. We rotate the target qubit to the $(\ket{\downarrow}+\ket{\uparrow})/\sqrt{2}$ state with a $\pi/2$ pulse. Then we apply the target gate of the sequence to accumulate phase error. We rotate the target qubit again to $\ket{\uparrow}$ using a $\pi/2$ pulse with an offset $\theta$ in phase. Finally, we measure the spin-up probability P$_{\uparrow}$ of the target qubit at the end of the sequence. 
We fit the sinusoidal modulation of P$_{\uparrow}$ as a function of $\theta$ to get the phase shift associated with each calibration sequence as shown in Fig.~\ref{fig2} \rev{(c)}. The fitted phase shifts are $(\theta_{\text{ZCROT}_{12}}, \theta_{\text{CROT}_{12}}, \theta_{\text{ZCROT}_{21}}, \theta_{\text{CROT}_{21}})= (0.41\pm 0.025, -0.58\pm 0.02, 0.43\pm 0.015, -0.49 \pm 0.03)~\text{rad}$.

The Hamiltonian in Eq.~(\ref{eq:H_rot}) implies that when a CROT$_{12}$ gate is applied, the $e^{\pm 2i \pi Jt}$ terms result in phases accumulated in the $\ket{\downarrow\downarrow}$ and $\ket{\downarrow\uparrow}$ states. From the full CROT Hamiltonian it follows that phase is always accumulated in the states which the pulse is not acting on (see Methods~\ref{methods:sim}). Fig.~\ref{fig2} \rev{(d)} shows the relation between the four pulses and their corresponding off-resonant Hamiltonian phase errors. 
To obtain the relation between the fitted phase shifts and the off-resonant Hamiltonian phase errors, we write down a table of phase errors accumulated on the four basis states along the sequence and obtain a relation between fitted phase shifts and the off-resonant Hamiltonian phase errors (see Methods~\ref{methods:calibration_seq}). Using this relation and the phase shifts measured, we obtain the off-resonant Hamiltonian phase errors $(\phi_{1,\downarrow},\phi_{1,\uparrow},\phi_{2,\downarrow},\phi_{2,\uparrow})=(-0.07\pm 0.03, 0.14\pm 0.02, -0.07\pm 0.03, 0.12\pm 0.03)~\text{rad}$. We note that the phase errors are time-independent coherent errors and therefore we can cancel the effect by characterizing the accumulated phases.

We use the measured off-resonant Hamiltonian phase errors to compensate for the unwanted phases accumulated in the calibration sequences. Fig.~\ref{fig2} \rev{(e)} shows the procedure we use to compensate for the phase errors in the CROT$_{12}$ sequence. We record all the phase errors at each step of the sequence. From these phase errors, we obtain the offset needed for each pulse to compensate for the effect of phase error, which is the phase accumulated on the states the pulse is acting on (see Methods~\ref{methods:calibration_seq}). 
To compensate for the accumulated phase error, we subtract the accumulated phase from the microwave phase, which implements a virtual z-gate \cite{mckay_efficient_2017}. These virtual z-gates change the rotating frame according to the accumulated phase errors such that the effects on the qubit gates caused by the phase errors are canceled.
Fig.~\ref{fig2} \rev{(f)} shows the measured curves after compensating phase errors in the sequences. The fitted phase shifts after compensation are $(\theta^{'}_{\text{ZCROT}_{12}}, \theta^{'}_{\text{CROT}_{12}}, \theta^{'}_{\text{ZCROT}_{21}}, \theta^{'}_{\text{CROT}_{21}}) = (0.15\pm 0.03, -0.02\pm 0.03, 0.08\pm 0.04, 0.01\pm 0.04)~\text{rad}$. After the phase compensation, all the phase shifts are reduced to $\leq 0.15~\text{rad}$ $\sim 9^{\circ}$. We notice that there are still non-zero phase shifts in the two ZCROT sequences. This is also observed in simulation (see Methods~\ref{methods:sim} and Supplementary Fig.~S3) from which we notice that this phase is not originating from the far off-resonant terms. This residual phase is only observed in sequences when both qubits are operated and its origin is not yet clear. A possible source for this residual phase could be a \rev{correlated gate error} on $Q_{2}$ ($Q_{1}$) when applying gates acting on $Q_{1}$ ($Q_{2}$) \cite{xue_quantum_2022}.

\subsection{Compensation of the off-resonant Hamiltonian phase error}\label{result:compensation}
\begin{figure}[H]
\centering
\captionsetup{font=footnotesize,skip=0pt,width=1\linewidth}
\includegraphics[width=1\textwidth]{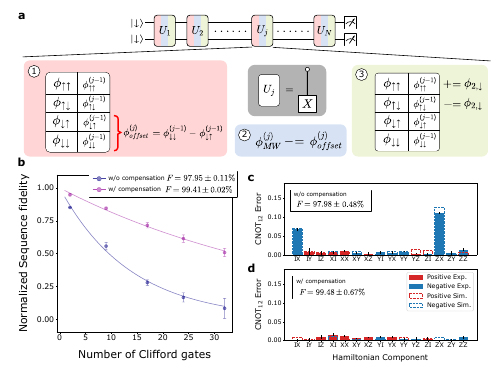}
\caption{\textbf{Off-resonant Hamiltonian phase error compensation procedure and performance.} 
\textbf{a} General off-resonant Hamiltonian phase error compensation procedure. We check the phase error table for phases accumulated in the relevant states for each pulse \rev{(1)}. We then subtract the offset from the microwave phase of the pulse \rev{(2)}. Finally, the phase error caused by the pulse is added to the phase error table and used to obtain the offsets for the following pulses \rev{(3)}.
\textbf{b} Results of two-qubit randomized benchmarking with and without the compensation procedure. The RB without the compensation procedure shows an averaged primitive gate fidelity of $97.95 \pm 0.11 \%$. The compensation procedure increases this fidelity to $99.41\pm 0.02 \%$. 
\textbf{c. d} Coefficients of the CNOT$_{12}$ gate error generator decomposition with and without compensation. The solid bars in the plot represent the experimental result, while the dashed bars represent the simulations. Large IX and ZX Hamiltonian component occurs when the compensation procedure is not applied. The two dominant error terms in (c) are significantly reduced when applying the compensation procedure. Experiment and simulation results show good consistency. The error bars represent $1\sigma$ \lcc{standard deviation} from the mean.}\label{fig3}
\end{figure}

Next, we extend the use of the compensation procedure to general pulse sequences. Fig.~\ref{fig3} (a) shows the phase compensation procedure for general pulse sequence \rev{$U_{1}, U_{2}, ..., U_{N}$}. We keep a phase error table that records the phase errors accumulated on the four basis states and check the sequence pulse by pulse with this phase error table to obtain the phase offsets. \rev{Since the errors are time-idependent, we can compensate the phase errors pulse by pulse.} For each applied pulse, we check the phase error which is accumulated on the states which the pulse acts on and offset the microwave phase correspondingly as done in the previous section. We then add the phase error accumulated by the pulse to the phase error table and move on to the next pulse. This procedure is performed in software before the execution of the physical pulses. We emphasize that this method can also be implemented in real-time, e.g., on an FPGA using a phase counter \cite{philips_universal_2022}.

To evaluate the performance of our calibration, we compare the pulse fidelity with and without the compensation by performing a two-qubit RB experiment \cite{noiri_fast_2022,huang_fidelity_2019}. We use 15 (59) different random sequences for the experiment without (with) the phase compensation protocol (see Methods~\ref{methods:2qrb}). Fig.~\ref{fig3} (b) shows measured RB decays. In the run without the compensation procedure, a Clifford fidelity of $94.73\pm 0.28\%$ is obtained, corresponding to a primitive gate fidelity of $97.95 \pm 0.11 \%$. With the compensation procedure, we achieve a Clifford fidelity of $98.48\pm 0.06\%$, corresponding to primitive gate fidelity of $99.41\pm 0.02 \%$. The increase in gate fidelity demonstrates that the compensation procedure has significantly reduced the errors in the pulses. 

To get a more detailed report on the performance of the quantum gates, we conduct a GST experiment \cite{nielsen_gate_2021, nielsen_probing_2020} (see also Methods~\ref{methods:gst}). Here the GST experiment generates the \rev{two-qubit} Pauli transformation matrix (PTM) of the implemented quantum gates, which describes how Pauli matrices are transformed under the quantum gate. The experimentally obtained PTM is then compared to the ideal PTM to get the error generator which \lcc{gives more specific gate error processes}. By writing the error generator into a linear combination of terms representing different error processes, we can interpret the errors of \lcc{our gates} more intuitively \cite{blume-kohout_taxonomy_2022}.

Fig.~\ref{fig3} (c) shows the error generator of the CNOT$_{12}$ gate obtained by both experiment and simulated GST. \lcc{The simulated data is obtained using} an ideal Hamiltonian \rev{(see Eq.~(\ref{eq:H_rot}))} without introducing any noise (see Methods~\ref{methods:sim}). Without the compensation, there are large errors in the IX and ZX Hamiltonian elements, both in simulation and experimental results. The consistency between the experiment and simulation shows the off-resonant Hamiltonian phase error considered in the Hamiltonian given in Eq.~(\ref{eq:H_rot}) is indeed the \lcc{dominant} error we measured in the experiment. The two terms are significantly suppressed when using the compensation procedure, as shown in Fig.~\ref{fig3} (d). Both experiment and simulation results exhibit this suppression, showing that the off-resonant Hamiltonian phase errors are \lcc{understood and} corrected as expected. The full GST gate metrics of the experiment with the phase error compensation protocol are shown in Supplementary Table~S1.

After the phase compensation protocol, the CNOT$_{12}$ gate still has an infidelity of $\sim 0.5$~\%. The precision of our GST result does not allow us to make a definite statement on whether the source of this infidelity is coherent Hamiltonian errors or incoherent stochastic errors. The \rev{correlated gate error} mentioned in Sec.~\ref{result:measuring} could be an indication that there are still Hamiltonian errors in our gate. Further investigations are required to determine whether the CNOT$_{12}$ fidelity can be increased by compensating this error.

\subsection{Virtual CZ Gate}\label{result:virtual}
\begin{figure}[H]
\centering
\captionsetup{font=footnotesize,skip=0pt,width=1\linewidth}
\includegraphics[width=0.75\textwidth]{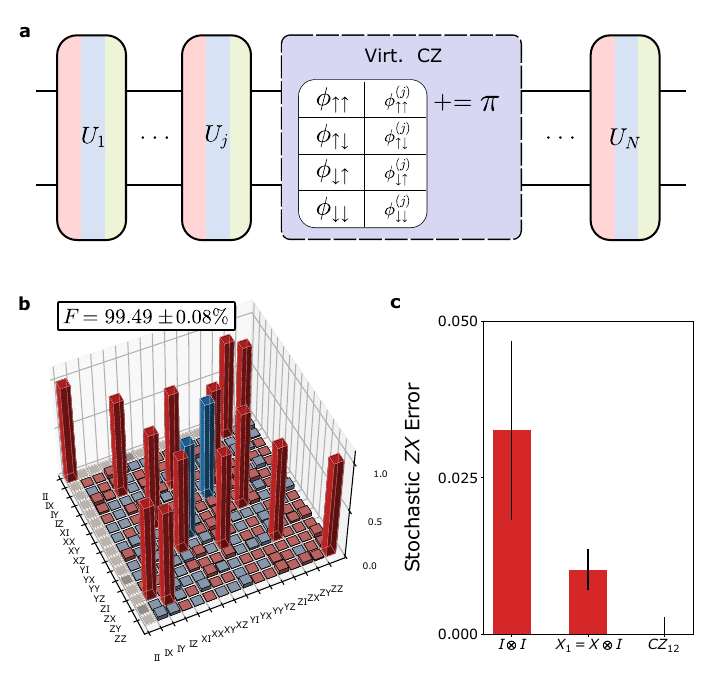}
\caption{\textbf{Virtual CZ} 
\textbf{a} Implementation of virtual CZ$_{12}$. A virtual CZ$_{12}$ is \lcc{executed} by adding a $\pi$ phase to the $\ket{\uparrow\uparrow}$ state in the table of phase error. Other controlled $z$-gates CZ$_{ij}$ can be implemented \lcc{analogously}.
\textbf{b} Gate-set tomography result of the virtual CZ$_{12}$ \lcc{showing the PTM obtained with experimental (solid line bars) and simulated (dashed line bars) datasets.}
\textbf{c} Comparison of stochastic noise in different gates. The graph shows the dominant stochastic error generator component ZX for the \rev{$I\otimes I$, $X_{1}=X\otimes I$}, and virtual CZ$_{12}$. We implement the identity gate by idling both qubits for the duration of a $\pi/2$ gate time (62 ns), and $X_{1}$ by applying a CROT$_{21}$ and a ZCROT$_{21}$ sequentially. Error bars represent the $1\sigma$ standard deviation from the mean.}\label{fig4}
\end{figure}

Finally, we demonstrate the implementation of a virtual CZ gate with the compensation procedure, shown in Fig.~\ref{fig4} (a). In the compensation procedure, we use the phase error table to obtain the microwave phase offset for each pulse in the sequence \rev{$U_{1}, U_{2}, ..., U_{N}$}. When a $\pi$ phase is added to the $\ket{\uparrow\uparrow}$ \lcc{row}, the following CROT$_{12}$ pulses will acquire an additional $\pi$ phase in the offset while the offsets of ZCROT$_{12}$ pulses are unchanged. This results in a $\pi$ phase difference between CROT$_{12}$ and ZCROT$_{12}$ pulses which is effectively equivalent to a CZ$_{12}$ gate.

To verify the virtual CZ$_{12}$, we perform a GST experiment with the CZ$_{12}$ gate-set. Fig.~\ref{fig4} (b) shows the estimated PTM of the virtual CZ$_{12}$. The measured PTM is close to the ideal CZ$_{12}$ and has a fidelity of $99.49\pm 0.08 \%$ which we anticipate can be increased by further phase calibrations. This shows that the virtual CZ$_{12}$ is well implemented. 

One advantage of virtually implementing quantum gates is that the gate time is reduced to zero such that the qubits are not affected by dephasing. Fig.~\ref{fig4} (c) shows the ZX stochastic error component for \rev{$I\otimes I$, $X_{1}=X\otimes I$} and virtual CZ$_{12}$. We use the identity gate \rev{$I\otimes I$} which idles both qubits for 62 ns to emulate the dephasing in a physical CZ gate. We find experimentally that ZX is the dominant component for these gates (see Supplementary Fig.~S5). There is a large ZX stochastic term for the identity gate due to dephasing by \rev{residual nuclear spin or charge noise} \cite{yoneda_noise-correlation_nodate}. This term is reduced in $X_{1}$ as the driven qubit is less affected by \rev{dephasing}  because the \lcc{drive effectively acts as a filter function} \cite{yoneda_quantum-dot_2018, Bylander2011, stano_review_2022, Laucht2016}. For the virtual CZ$_{12}$, this noise is reduced to essentially zero, showing that the virtual CZ$_{12}$ is indeed not affected by charge noise, as expected.

\section{Discussion}\label{dicussion}
Understanding the off-resonant Hamiltonian phase errors allows us to implement high-fidelity CROT gates systematically. 
With a better understanding of the origin of these phase errors, we can estimate the experimental phase shifts by numerical simulations (see Supplementary Fig.~S3).
The phase error compensation method also allows the implementation of a virtual CZ gate, which is useful in quantum circuits that use controlled-phase gates extensively. For example, the quantum Fourier transformation (QFT) circuit uses controlled-phase gates of several different rotation angles \cite{nielsen_quantum_2010}. \lcc{Our procedure is implementable on an FPGA which tracks and corrects the phase offsets in real-time. Such a real-time approach is crucial for quantum circuits that cannot be pre-calculated before execution, e.g., for quantum error correction \cite{takeda_quantum_2022} or real-time feedback initialization \cite{philips_universal_2022}.}

In the original proposal of the CROT gate \cite{russ_high-fidelity_2018,zajac_resonantly_2018}, the exchange coupling is only turned on for the execution of the CROT and turned off for subsequent single-qubit operations. The pulse to control the exchange effectively results in a CPhase gate up to single-qubit z-rotations \cite{russ_high-fidelity_2018}. This exchange coupling pulse length can be calibrated such that the CPhase gate accumulates a controlled-phase of one full rotation. A microwave pulse within this exchange pulse is applied to implement a CNOT gate up to single-qubit z-rotations \cite{russ_high-fidelity_2018}. In this case, we anticipate the off-resonant Hamiltonian phase errors discussed here can be canceled out by adjusting the exchange pulse time to control the accumulation of an extra controlled-phase.

The practical scalability of the exchange-always-on system remains an open question. For the three-qubit resonantly-driven Toffoli gate in the exchange-controlled system, the extra conditional phase error can be removed by changing the timing of the exchange coupling pulse \cite{takeda_quantum_2022, Gullans2019}. This exchange pulse accumulates additional, conditional phases and has the same effect as shifting the microwave phases. In a three-qubit exchange-always-on system, keeping track of all the transition frequencies becomes more challenging, and the calibration sequences for measuring the off-resonant Hamiltonian phase errors become more complicated. While further investigations are required, we anticipate that the procedure discussed here can also be used to calibrate the controlled rotations in a three-qubit exchange-always-on system.

In summary, we demonstrated a systematic way to calibrate high-fidelity CROT gates in an exchange-always-on \lcc{two-qubit} system. We present a calibration procedure to compensate for the Hamiltonian off-resonant phase errors in our CROT gates, allowing us to achieve universal single and two-qubit gate fidelities above the fault-tolerant threshold of 99\%. Finally, we implement a high-fidelity virtual CZ gate using our phase error compensation protocol.

\section*{Data Availability}
All data of this study will be made available in a Zenodo repository.

\section*{Acknowledgments}
Y.H.W. acknowledges useful discussions with C. Chiang.
This work was supported financially by Core Research for Evolutional Science and Technology (CREST), Japan Science and Technology Agency (JST)  (JPMJCR1675), MEXT Quantum Leap Flagship Program (MEXT Q-LEAP) grant numbers JPMXS0118069228, JST Moonshot R\&D grant number JPMJMS226B-1, and JSPS KAKENHI grant numbers 18H01819 and 20H00237. T.N. acknowledges support from JST PRESTO grant number JPMJPR2017. L.C.C. acknowledges support by a Swiss NSF mobility fellowship (P2BSP2\textunderscore 200127). Y.H.W. acknowledges support by RIKEN's IPA program and National Taiwan University Higher Education SPROUT Project Research Promotion Program for Direct-Entry Doctoral Degree Program Students (L4100). H.-S.G. acknowledges support from the National Science and Technology Council (NSTC), Taiwan under Grants No. NSTC 112-2119-M-002 -014, No. NSTC 111-2119-M-002-007, No. NSTC 111-2119-M-002-006-MY3, No. NSTC
111-2627-M-002-001, and No. NSTC 111-2622-8-002-001, and from the National Taiwan University under Grants No. NTU-CC-111L894604, and No. NTUCC-112L893404. H.-S.G. is grateful to the support from the Physics Division, National Center for Theoretical Sciences, Taiwan

\section*{Author Contribution}
Y.H.W. and L.C.C. performed the experiment. A.N. fabricated the device. K.T., A.N.,  T.K., T.N., C.Y.C. and H.S.G. contributed to the data acquisition and discussed the results. A.S. and G.S. developed and supplied the silicon-28/silicon-germanium heterostructure. Y.H.W. and L.C.C. wrote the manuscript with inputs from all co-authors. S.T. supervised the project.

\section*{Declarations}
The authors declare no competing interests.

\section*{Additional Information}
\textbf{Correspondence and requests for materials} should be addressed to Y.H.W. (email: yi-hsien.wu@riken.jp), L.C.C. (email: leon.camenzind@riken.jp) or S.T. (email: tarucha@riken.jp).

\bibliography{CROT_phase_correct_bib}

\begin{thebibliography}{10}
\expandafter\ifx\csname url\endcsname\relax
  \def\url#1{\burl{#1}}\fi
\expandafter\ifx\csname urlprefix\endcsname\relax\def\urlprefix{URL }\fi
\providecommand{\bibinfo}[2]{#2}
\providecommand{\eprint}[2][]{\url{#2}}
\providecommand{\doi}[1]{\url{https://doi.org/#1}}

\bibitem{loss_quantum_1998}
\bibinfo{author}{Loss, D.} \& \bibinfo{author}{DiVincenzo, D.~P.}
\newblock \bibinfo{title}{Quantum computation with quantum dots}.
\newblock \emph{\bibinfo{journal}{Physical Review A}}
  \textbf{\bibinfo{volume}{57}}~(1), \bibinfo{pages}{120--126}
  (\bibinfo{year}{1998}) .

\bibitem{philips_universal_2022}
\bibinfo{author}{Philips, S. G.~J.} \emph{et~al.}
\newblock \bibinfo{title}{Universal control of a six-qubit quantum processor in
  silicon}.
\newblock \emph{\bibinfo{journal}{Nature}}
  \textbf{\bibinfo{volume}{609}}~(7929), \bibinfo{pages}{919--924}
  (\bibinfo{year}{2022}) .

\bibitem{hendrickx_four-qubit_2021}
\bibinfo{author}{Hendrickx, N.~W.} \emph{et~al.}
\newblock \bibinfo{title}{A four-qubit germanium quantum processor}.
\newblock \emph{\bibinfo{journal}{Nature}}
  \textbf{\bibinfo{volume}{591}}~(7851), \bibinfo{pages}{580--585}
  (\bibinfo{year}{2021}) .

\bibitem{yoneda_quantum-dot_2018}
\bibinfo{author}{Yoneda, J.} \emph{et~al.}
\newblock \bibinfo{title}{A quantum-dot spin qubit with coherence limited by
  charge noise and fidelity higher than 99.9\%}.
\newblock \emph{\bibinfo{journal}{Nature Nanotechnology}}
  \textbf{\bibinfo{volume}{13}}~(2), \bibinfo{pages}{102--106}
  (\bibinfo{year}{2018}) .

\bibitem{veldhorst_addressable_2014}
\bibinfo{author}{Veldhorst, M.} \emph{et~al.}
\newblock \bibinfo{title}{An addressable quantum dot qubit with fault-tolerant
  control-fidelity}.
\newblock \emph{\bibinfo{journal}{Nature Nanotechnology}}
  \textbf{\bibinfo{volume}{9}}~(12), \bibinfo{pages}{981--985}
  (\bibinfo{year}{2014}) .

\bibitem{yoneda_fast_2014}
\bibinfo{author}{Yoneda, J.} \emph{et~al.}
\newblock \bibinfo{title}{Fast {Electrical} {Control} of {Single} {Electron}
  {Spins} in {Quantum} {Dots} with {Vanishing} {Influence} from {Nuclear}
  {Spins}}.
\newblock \emph{\bibinfo{journal}{Physical Review Letters}}
  \textbf{\bibinfo{volume}{113}}~(26), \bibinfo{pages}{267601}
  (\bibinfo{year}{2014}) .

\bibitem{yang_operation_2020}
\bibinfo{author}{Yang, C.~H.} \emph{et~al.}
\newblock \bibinfo{title}{Operation of a silicon quantum processor unit cell
  above one kelvin}.
\newblock \emph{\bibinfo{journal}{Nature}}
  \textbf{\bibinfo{volume}{580}}~(7803), \bibinfo{pages}{350--354}
  (\bibinfo{year}{2020}) .

\bibitem{petit_universal_2020}
\bibinfo{author}{Petit, L.} \emph{et~al.}
\newblock \bibinfo{title}{Universal quantum logic in hot silicon qubits}.
\newblock \emph{\bibinfo{journal}{Nature}}
  \textbf{\bibinfo{volume}{580}}~(7803), \bibinfo{pages}{355--359}
  (\bibinfo{year}{2020}) .

\bibitem{camenzind_spin_2022}
\bibinfo{author}{Camenzind, L.~C.} \emph{et~al.}
\newblock \bibinfo{title}{A spin qubit in a fin field-effect transistor}.
\newblock \emph{\bibinfo{journal}{Nature Electronics}}
  \textbf{\bibinfo{volume}{5}}~(3), \bibinfo{pages}{178--183}
  (\bibinfo{year}{2022}).
\newblock \bibinfo{note}{ArXiv:2103.07369 [cond-mat, physics:quant-ph]} .

\bibitem{zwerver_qubits_2022}
\bibinfo{author}{Zwerver, A. M.~J.} \emph{et~al.}
\newblock \bibinfo{title}{Qubits made by advanced semiconductor manufacturing}.
\newblock \emph{\bibinfo{journal}{Nature Electronics}}
  \textbf{\bibinfo{volume}{5}}~(3), \bibinfo{pages}{184--190}
  (\bibinfo{year}{2022}) .

\bibitem{xue_cmos-based_2021}
\bibinfo{author}{Xue, X.} \emph{et~al.}
\newblock \bibinfo{title}{{CMOS}-based cryogenic control of silicon quantum
  circuits}.
\newblock \emph{\bibinfo{journal}{Nature}}
  \textbf{\bibinfo{volume}{593}}~(7858), \bibinfo{pages}{205--210}
  (\bibinfo{year}{2021}) .

\bibitem{Vandersypen_interfacing_2017}
\bibinfo{author}{Vandersypen, L. M.~K.} \emph{et~al.}
\newblock \bibinfo{title}{Interfacing spin qubits in quantum dots and
  donors{\textemdash}hot, dense, and coherent}.
\newblock \emph{\bibinfo{journal}{npj Quantum Information}}
  \textbf{\bibinfo{volume}{3}}~(1) (\bibinfo{year}{2017}) .

\bibitem{takeda_quantum_2022}
\bibinfo{author}{Takeda, K.}, \bibinfo{author}{Noiri, A.},
  \bibinfo{author}{Nakajima, T.}, \bibinfo{author}{Kobayashi, T.} \&
  \bibinfo{author}{Tarucha, S.}
\newblock \bibinfo{title}{Quantum error correction with silicon spin qubits}.
\newblock \emph{\bibinfo{journal}{Nature}}
  \textbf{\bibinfo{volume}{608}}~(7924), \bibinfo{pages}{682--686}
  (\bibinfo{year}{2022}) .

\bibitem{fowler_surface_2012}
\bibinfo{author}{Fowler, A.~G.}, \bibinfo{author}{Mariantoni, M.},
  \bibinfo{author}{Martinis, J.~M.} \& \bibinfo{author}{Cleland, A.~N.}
\newblock \bibinfo{title}{Surface codes: {Towards} practical large-scale
  quantum computation}.
\newblock \emph{\bibinfo{journal}{Physical Review A}}
  \textbf{\bibinfo{volume}{86}}~(3), \bibinfo{pages}{032324}
  (\bibinfo{year}{2012}) .

\bibitem{wang_surface_2011}
\bibinfo{author}{Wang, D.~S.}, \bibinfo{author}{Fowler, A.~G.} \&
  \bibinfo{author}{Hollenberg, L. C.~L.}
\newblock \bibinfo{title}{Surface code quantum computing with error rates over
  1\%}.
\newblock \emph{\bibinfo{journal}{Physical Review A}}
  \textbf{\bibinfo{volume}{83}}~(2), \bibinfo{pages}{020302}
  (\bibinfo{year}{2011}) .

\bibitem{yang_silicon_2019}
\bibinfo{author}{Yang, C.~H.} \emph{et~al.}
\newblock \bibinfo{title}{Silicon qubit fidelities approaching incoherent noise
  limits via pulse engineering}.
\newblock \emph{\bibinfo{journal}{Nature Electronics}}
  \textbf{\bibinfo{volume}{2}}~(4), \bibinfo{pages}{151--158}
  (\bibinfo{year}{2019}) .

\bibitem{noiri_fast_2022}
\bibinfo{author}{Noiri, A.} \emph{et~al.}
\newblock \bibinfo{title}{Fast universal quantum gate above the fault-tolerance
  threshold in silicon}.
\newblock \emph{\bibinfo{journal}{Nature}}
  \textbf{\bibinfo{volume}{601}}~(7893), \bibinfo{pages}{338--342}
  (\bibinfo{year}{2022}) .

\bibitem{xue_quantum_2022}
\bibinfo{author}{Xue, X.} \emph{et~al.}
\newblock \bibinfo{title}{Quantum logic with spin qubits crossing the surface
  code threshold}.
\newblock \emph{\bibinfo{journal}{Nature}}
  \textbf{\bibinfo{volume}{601}}~(7893), \bibinfo{pages}{343--347}
  (\bibinfo{year}{2022}) .

\bibitem{mills_two-qubit_2022}
\bibinfo{author}{Mills, A.~R.} \emph{et~al.}
\newblock \bibinfo{title}{Two-qubit silicon quantum processor with operation
  fidelity exceeding 99\%}.
\newblock \emph{\bibinfo{journal}{Science Advances}}
  \textbf{\bibinfo{volume}{8}}~(14), \bibinfo{pages}{eabn5130}
  (\bibinfo{year}{2022}) .

\bibitem{huang_fidelity_2019}
\bibinfo{author}{Huang, W.} \emph{et~al.}
\newblock \bibinfo{title}{Fidelity benchmarks for two-qubit gates in silicon}.
\newblock \emph{\bibinfo{journal}{Nature}}
  \textbf{\bibinfo{volume}{569}}~(7757), \bibinfo{pages}{532--536}
  (\bibinfo{year}{2019}) .

\bibitem{mckay_efficient_2017}
\bibinfo{author}{McKay, D.~C.}, \bibinfo{author}{Wood, C.~J.},
  \bibinfo{author}{Sheldon, S.}, \bibinfo{author}{Chow, J.~M.} \&
  \bibinfo{author}{Gambetta, J.~M.}
\newblock \bibinfo{title}{Efficient {Z} gates for quantum computing}.
\newblock \emph{\bibinfo{journal}{Physical Review A}}
  \textbf{\bibinfo{volume}{96}}~(2), \bibinfo{pages}{022330}
  (\bibinfo{year}{2017}) .

\bibitem{magesan_scalable_2011}
\bibinfo{author}{Magesan, E.}, \bibinfo{author}{Gambetta, J.~M.} \&
  \bibinfo{author}{Emerson, J.}
\newblock \bibinfo{title}{Scalable and {Robust} {Randomized} {Benchmarking} of
  {Quantum} {Processes}}.
\newblock \emph{\bibinfo{journal}{Physical Review Letters}}
  \textbf{\bibinfo{volume}{106}}~(18), \bibinfo{pages}{180504}
  (\bibinfo{year}{2011}) .

\bibitem{magesan_efficient_2012}
\bibinfo{author}{Magesan, E.} \emph{et~al.}
\newblock \bibinfo{title}{Efficient {Measurement} of {Quantum} {Gate} {Error}
  by {Interleaved} {Randomized} {Benchmarking}}.
\newblock \emph{\bibinfo{journal}{Physical Review Letters}}
  \textbf{\bibinfo{volume}{109}}~(8), \bibinfo{pages}{080505}
  (\bibinfo{year}{2012}) .

\bibitem{nielsen_gate_2021}
\bibinfo{author}{Nielsen, E.} \emph{et~al.}
\newblock \bibinfo{title}{Gate {Set} {Tomography}}.
\newblock \emph{\bibinfo{journal}{Quantum}} \textbf{\bibinfo{volume}{5}},
  \bibinfo{pages}{557} (\bibinfo{year}{2021}).
\newblock \bibinfo{note}{ArXiv:2009.07301 [quant-ph]} .

\bibitem{noiri_radio-frequency-detected_2020}
\bibinfo{author}{Noiri, A.} \emph{et~al.}
\newblock \bibinfo{title}{Radio-{Frequency}-{Detected} {Fast} {Charge}
  {Sensing} in {Undoped} {Silicon} {Quantum} {Dots}}.
\newblock \emph{\bibinfo{journal}{Nano Letters}}
  \textbf{\bibinfo{volume}{20}}~(2), \bibinfo{pages}{947--952}
  (\bibinfo{year}{2020}) .

\bibitem{elzerman_single-shot_2004}
\bibinfo{author}{Elzerman, J.~M.} \emph{et~al.}
\newblock \bibinfo{title}{Single-shot read-out of an individual electron spin
  in a quantum dot}.
\newblock \emph{\bibinfo{journal}{Nature}}
  \textbf{\bibinfo{volume}{430}}~(6998), \bibinfo{pages}{431--435}
  (\bibinfo{year}{2004}) .

\bibitem{russ_high-fidelity_2018}
\bibinfo{author}{Russ, M.} \emph{et~al.}
\newblock \bibinfo{title}{High-fidelity quantum gates in {Si}/{SiGe} double
  quantum dots}.
\newblock \emph{\bibinfo{journal}{Physical Review B}}
  \textbf{\bibinfo{volume}{97}}~(8), \bibinfo{pages}{085421}
  (\bibinfo{year}{2018}) .

\bibitem{zajac_resonantly_2018}
\bibinfo{author}{Zajac, D.~M.} \emph{et~al.}
\newblock \bibinfo{title}{Resonantly driven {CNOT} gate for electron spins}.
\newblock \emph{\bibinfo{journal}{Science}}
  \textbf{\bibinfo{volume}{359}}~(6374), \bibinfo{pages}{439--442}
  (\bibinfo{year}{2018}) .

\bibitem{nielsen_probing_2020}
\bibinfo{author}{Nielsen, E.} \emph{et~al.}
\newblock \bibinfo{title}{Probing quantum processor performance with {pyGSTi}}.
\newblock \emph{\bibinfo{journal}{Quantum Science and Technology}}
  \textbf{\bibinfo{volume}{5}}~(4), \bibinfo{pages}{044002}
  (\bibinfo{year}{2020}) .

\bibitem{blume-kohout_taxonomy_2022}
\bibinfo{author}{Blume-Kohout, R.} \emph{et~al.}
\newblock \bibinfo{title}{A {Taxonomy} of {Small} {Markovian} {Errors}}.
\newblock \emph{\bibinfo{journal}{PRX Quantum}}
  \textbf{\bibinfo{volume}{3}}~(2), \bibinfo{pages}{020335}
  (\bibinfo{year}{2022}) .

\bibitem{yoneda_noise-correlation_nodate}
\bibinfo{author}{Yoneda, J.} \emph{et~al.}
\newblock \bibinfo{title}{Noise-correlation spectrum for a pair of spin qubits
  in silicon}  (\bibinfo{year}{2022}).
\newblock \bibinfo{note}{ArXiv:2208.14150 [quant-ph]} .

\bibitem{Bylander2011}
\bibinfo{author}{Bylander, J.} \emph{et~al.}
\newblock \bibinfo{title}{Noise spectroscopy through dynamical decoupling with
  a superconducting flux qubit}.
\newblock \emph{\bibinfo{journal}{Nature Physics}}
  \textbf{\bibinfo{volume}{7}}~(7), \bibinfo{pages}{565--570}
  (\bibinfo{year}{2011}) .

\bibitem{stano_review_2022}
\bibinfo{author}{Stano, P.} \& \bibinfo{author}{Loss, D.}
\newblock \bibinfo{title}{Review of performance metrics of spin qubits in gated
  semiconducting nanostructures}.
\newblock \emph{\bibinfo{journal}{Nature Reviews Physics}}
  \textbf{\bibinfo{volume}{4}}~(10), \bibinfo{pages}{672--688}
  (\bibinfo{year}{2022}) .

\bibitem{Laucht2016}
\bibinfo{author}{Laucht, A.} \emph{et~al.}
\newblock \bibinfo{title}{A dressed spin qubit in silicon}.
\newblock \emph{\bibinfo{journal}{Nature Nanotechnology}}
  \textbf{\bibinfo{volume}{12}}~(1), \bibinfo{pages}{61--66}
  (\bibinfo{year}{2016}) .

\bibitem{nielsen_quantum_2010}
\bibinfo{author}{Nielsen, M.~A.} \& \bibinfo{author}{Chuang, I.~L.}
\newblock \emph{\bibinfo{title}{Quantum computation and quantum information}}
  \bibinfo{edition}{10th anniversary ed} edn (\bibinfo{publisher}{Cambridge
  University Press}, \bibinfo{address}{Cambridge ; New York},
  \bibinfo{year}{2010}).

\bibitem{Gullans2019}
\bibinfo{author}{Gullans, M.~J.} \& \bibinfo{author}{Petta, J.~R.}
\newblock \bibinfo{title}{Protocol for a resonantly driven three-qubit toffoli
  gate with silicon spin qubits}.
\newblock \emph{\bibinfo{journal}{Physical Review B}}
  \textbf{\bibinfo{volume}{100}}~(8) (\bibinfo{year}{2019}) .

\bibitem{barends_superconducting_2014}
\bibinfo{author}{Barends, R.} \emph{et~al.}
\newblock \bibinfo{title}{Superconducting quantum circuits at the surface code
  threshold for fault tolerance}.
\newblock \emph{\bibinfo{journal}{Nature}}
  \textbf{\bibinfo{volume}{508}}~(7497), \bibinfo{pages}{500--503}
  (\bibinfo{year}{2014}) .

\bibitem{madzik_precision_2022}
\bibinfo{author}{Mądzik, M.~T.} \emph{et~al.}
\newblock \bibinfo{title}{Precision tomography of a three-qubit donor quantum
  processor in silicon}.
\newblock \emph{\bibinfo{journal}{Nature}}
  \textbf{\bibinfo{volume}{601}}~(7893), \bibinfo{pages}{348--353}
  (\bibinfo{year}{2022}) .

\end{thebibliography}
\bibliographystyle{sn-standardnature_nourl}

\section{Methods}\label{methods}

\subsection{Calibration Sequences} \label{methods:calibration_seq}
Supplementary Fig.~S1 shows all four calibration sequences and the corresponding phase error table used to obtain the phase offsets. For the ZCROT$_{12}$ sequence, the phase errors associated with the four states at the end of the sequence are
\begin{align}
\phi_{\uparrow\uparrow} &= 2\phi_{1,\downarrow} + 2\phi_{2,\downarrow} , \\
\phi_{\uparrow\downarrow} &= -2\phi_{1,\downarrow}, \\
\phi_{\downarrow\uparrow} &= 2\phi_{1,\uparrow} -2\phi_{2,\downarrow}, \\
\phi_{\downarrow\downarrow} &= -2\phi_{1,\uparrow}.
\end{align}
The Ramsey sequence at the end measures the relative phase between the two off-resonant states of ZCROT$_{12}$, the $\ket{\uparrow\downarrow}$ and $\ket{\uparrow\uparrow}$ state. Thus the measured phase shift is 
\rev{\begin{equation}
\theta_{\text{ZCROT}_{12}} = \phi_{\uparrow\downarrow} - \phi_{\uparrow\uparrow} = - 4\phi_{1,\downarrow} - 2\phi_{2,\downarrow}.
\end{equation}}
Using a similar argument, we can write the phase shifts measured into a linear combination of the off-resonant Hamiltonian phase errors
\begin{align}
\theta_{\text{ZCROT}_{12}} &= -4\phi_{1,\downarrow} - 2\phi_{2,\downarrow},  \\
\theta_{\text{CROT}_{12}} &= -4\phi_{1,\uparrow}, \\
\theta_{\text{ZCROT}_{21}} &= -2\phi_{1,\downarrow} -4\phi_{2,\downarrow}, \\
\theta_{\text{CROT}_{21}} &= -4\phi_{2,\uparrow}.
\end{align}
Solving these four equations gives the explicit form of the off-resonant Hamiltonian phase errors. The relation is
\begin{align}
&(\phi_{1,\downarrow},\phi_{1,\uparrow},\phi_{2,\downarrow},\phi_{2,\uparrow}) \\
&= \left( \frac{-2\theta_{\text{ZCROT}_{12}} + \theta_{\text{ZCROT}_{21}}}{6}, -\frac{\theta_{\text{CROT}_{12}}}{4}, \frac{\theta_{\text{ZCROT}_{12}}-2\theta_{\text{ZCROT}_{21}}}{6}, -\frac{\theta_{\text{CROT}_{21}}}{4} \right) .
\end{align}
With the obtained off-resonant Hamiltonian phase errors we write down a table of phase errors accumulated on each basis states before the pulse is applied. We calculate the offsets needed by
\begin{align}
\phi_{offset, ZCROT_{12}} &= \phi_{\downarrow\downarrow}-\phi_{\downarrow\uparrow}, \\
\phi_{offset, CROT_{12}} &= \phi_{\uparrow\downarrow} - \phi_{\uparrow\uparrow}, \\
\phi_{offset, ZCROT_{21}} &= \phi_{\downarrow\downarrow} - \phi_{\uparrow\downarrow}, \\
\phi_{offset, CROT_{21}} &= \phi_{\downarrow\uparrow} - \phi_{\uparrow\uparrow}.
\end{align}

\subsection{Two-Qubit Randomized Benchmarking}\label{methods:2qrb}
We choose sequence lengths $L=(1,8,16,23,31)$ in our two-qubit randomized benchmarking experiment. For each length we use 59 (15) different random sequences for the experiment with (without) the phase compensation protocol. We combined three datasets measured over a time span of one week, demonstrating the stability of our qubits. For each sequence, the probability is obtained by averaging 150 single-shot measurements. The gates in each sequence are randomly chosen from the two-qubit Clifford group, which contains 11520 elements \cite{barends_superconducting_2014}. We use a computer search to find the combinations of primitive gates (see Supplementary Fig.~S2) to construct all two-qubit Clifford group elements \cite{noiri_fast_2022, huang_fidelity_2019}. At the end of the sequence, we search recovery gate, which projects the state into the target state $\ket{\uparrow\uparrow}$ or $\ket{\downarrow\downarrow}$. This results in two sequence fidelities $F_{\uparrow\uparrow}(n)$ and $F_{\downarrow\downarrow}(n)$. We fit the difference between two sequences $F(n)=F_{\uparrow\uparrow}(n)-F_{\downarrow\downarrow}(n)$ with the formula $F(n)=(A_{t}-B_{t})p^{n}$ where $A_{t}$ and $B_{t}$ absorbs the SPAM error and $p$ is the depolarizing strength. The two-qubit Clifford gate fidelity is obtained by $F_{C}=(1+3p)/4$. Each Clifford element is composed of 2.57 primitive gates on average. We therefore calculate the primitive gate fidelity as $F_{p} = 1 - (1-F_{C})/2.57$.

\subsection{CROT Simulation} \label{methods:sim}
We start with the Hamiltonian given in Eq.~(\ref{eq:H_lab}) and transform the Hamiltonian into the rotating frame using
\begin{align}
H_{R}(t) = R H R^{\dagger} - \frac{ih}{2\pi} \frac{\partial R}{\partial t} R^{\dagger},
\end{align}
with $R = \text{diag}(e^{-2i\pi E_{\text{z}} t}, e^{-i\pi(-\delta\tilde{E}_{\text{z}}-J)t}, e^{-i\pi(\delta\tilde{E}_{\text{z}}-J)t}, e^{2i\pi  E_{\text{z}} t} )$. The Hamiltonian in the rotating frame is then
\begin{equation}
H_{R}(t) = \frac{h}{2}\begin{pmatrix}
0 & B(t)e^{-2i\pi f_{2,\uparrow}t} & B(t)e^{-2i\pi f_{1,\uparrow}t} & 0 \\
B^{*}(t)e^{2i\pi f_{2,\uparrow} t }  & 0 & 0 & B(t)e^{-2i\pi f_{1,\downarrow} t } \\
B^{*}(t)e^{2i\pi f_{1,\uparrow}t } & 0 & 0 & B(t)e^{-2i\pi f_{2,\downarrow}t} \\
0 & B^{*}(t)e^{2i\pi f_{1,\downarrow} t } & B^{*}(t)e^{2i\pi f_{2,\downarrow}t } & 0
\end{pmatrix} ,
\end{equation}
with the effective EDSR magnetic field $B(t) = f_{R} e^{-2i\pi f_{\text{MW}} t + i\phi}$. We substitute this magnetic field into the rotating frame Hamiltonian and calculate the propagator. If the RWA is used, we set the elements in the far off-resonant terms to zero before calculating the propagator.

We choose $J=16$ MHz and $f_{R} = J/\sqrt{15} \simeq 4.13$ MHz, which results in a $\pi/2$ gate time $T_{\pi/2}\simeq 60.5$~ns. We compute the unitary propagator with this Hamiltonian by
\begin{align}
U(f_{\text{MW}}, \phi) = \mathcal{T} \exp \left( -\frac{i}{\hbar} \int_{0}^{T_{\pi/2}} H(f_{\text{MW}}, \phi, t) dt \right),
\end{align}
with $\mathcal{T}$ the time-ordering operator. By choosing the driving frequency $f_{\text{MW}}$, we select which ZCROT or CROT is implemented. Changing the microwave phase $\phi$ changes the rotation angle of the pulse. This unitary is then used for simulating the implemented pulses.

\subsection{Gate-set Tomography} \label{methods:gst}
For the GST experiments, depending on the implemented gates in the system, a different target gate set is chosen. This target gate set is then used to compose a preparation and measurement gate set and a set of germ sequences. The preparation- and measurement-fiducial gates are used to make tomographic measurements. These fiducial gates must be able to prepare and measure an information-complete set of states. The germ sequence in between is chosen from the germ set, which is amplificationally-complete \cite{nielsen_gate_2021} and therefore capable of amplifying all possible errors that can occur during the gate operation. 

We perform GST with the python package pyGSTi \cite{nielsen_probing_2020}. We use the default gate set provided by the pyGSTi package and the fiducial pair reduction function to reduce the number of sequences required. 
The CNOT$_{12}$ gate-set contains $\{ I, X_{1}, Y_{1}, X_{2}, Y_{2}, CNOT_{12}\}$. The identity gate $I$ is implemented by idling both qubits for a time of $T_{\pi/2} = 62$~ns. $X_{1,2}$ ($Y_{1,2}$) are $\pi/2$ rotations along the x-axis (y-axis) for $Q_{1}$ or $Q_{2}$, respectively. The 15 germs for this gate set are
\begin{align*}
&\{ I, X_{1}, Y_{1},X_{2},Y_{2}, CNOT_{12}, X_{1}Y_{1}, X_{2}Y_{2}, X_{1}X_{1}X_{1}, X_{2}X_{2}X_{2}, \\
&CNOT_{12}X_{2}X_{1}X_{1}, X_{1}X_{2}Y_{2}X_{1}Y_{2}Y_{1}, X_{1}Y_{2}X_{2}Y_{1}X_{2}X_{2}, \\
&Y_{1}Y_{2}X_{1}Y_{1}X_{1}CNOT_{12}, Y_{1}X_{2}Y_{2}X_{1}X_{2}X_{1}Y_{1}Y_{2} \},
\end{align*}
and the fiducial gates
\begin{align*}
&\{ \text{null}, X_{2}, Y_{2}, X_{2}X_{2}, X_{1}, X_{1}X_{2}, \\
&X_{1}Y_{2}, X_{1}X_{2}X_{2}, Y_{1}, Y_{1}X_{2}, Y_{1}Y_{2}, Y_{1}X_{2}X_{2}, X_{1}X_{1}, X_{1}X_{1}X_{2}, \\
&X_{1}X_{1}Y_{2}, X_{1}X_{1}X_{2}X_{2} \}.
\end{align*}
In contrast to the identity gate $I$, the $\text{null}$-gate has no physical idling time. We choose the sequence lengths $L=(1,2,4,8,16)$, which results in a total of 1760 sequences. 

For the CZ$_{12}$ gate-set which contains $\{ I, X_{1}, Y_{1}, X_{2}, Y_{2}, CZ_{12}\}$, the germs and \lcc{fiducial gates} are
\begin{align*}
&\{ I, X_{1}, Y_{1}, X_{2}, Y_{2}, CZ_{12}, X_{1}Y_{1} X_{2}Y_{2}, X_{1}X_{1}Y_{1}, \\
& X_{2}Y_{2}CZ_{12}, CZ_{12}X_{2}X_{1}X_{1}, X_{1}X_{2}Y_{2}X_{1}Y_{2}Y_{1}, X_{1}Y_{2}X_{2}Y_{1}X_{2}X_{2}, \\
&  CZ_{12}X_{2}Y_{1}CZ_{12}Y_{2}X_{1}, Y_{1}X_{1}Y_{2}X_{1}X_{2}X_{1}Y_{1}Y_{2} \},
\end{align*}
and
\begin{align*}
&\{ \text{null}, X_{2}, Y_{2}, X_{2}X_{2}, X_{1}, X_{1}X_{2}, \\
&X_{1}Y_{2}, X_{1}X_{2}X_{2}, Y_{1}, Y_{1}X_{2}, Y_{1}Y_{2}, Y_{1}X_{2}X_{2}, X_{1}X_{1}, X_{1}X_{1}X_{2}, \\
& X_{1}X_{1}Y_{2}, X_{1}X_{1}X_{2}X_{2} \},
\end{align*} 
which results in a total of 1644 sequences. All the gates used in the GST experiment are composed of CROT and ZCROT $\pi/2$ pulses and single-qubit z-rotations (see Supplementary Fig.~S4). The sequences are executed on the device to gather outcome counts. After execution of these sequences, the measured spin-up and spin-down counts are analyzed with an H+S model (see Methods~\ref{methods:error_generators}) to obtain the PTMs $G_{\text{exp}}$ of the gates. \lcc{This PTM} has the form
\begin{equation}
(G_{\text{exp}})_{ij} = \frac{1}{d}\tr[P_{i}G(P_{j})],
\end{equation}
where $d$ is the Hilbert space dimension and $P_{i}$ are the two-qubit Pauli operators. The estimated PTMs are then compared with the ideal PTMs to obtain the error generators through functions in the pyGSTi package.

To verify the assumption we made for the off-resonant Hamiltonian phase error, we also perform GST with simulated data sets. First, we take the sequences used in the experiment and calculate the corresponding series of unitary propagators with the ideal Hamiltonian. Then, we evolve the input ground state with this series of unitaries to obtain the final output state and calculate probabilities in each outcome to generate simulated counts. Finally, the simulated counts are analyzed in the same way as the experimental counts.

\subsection{Error Generators}\label{methods:error_generators}
For a noisy implementation $G_{\text{exp}}$ of the ideal quantum gate $G_{\text{ideal}}$, we can model the imperfect gate as 
\begin{align}
G_{\text{exp}} = \mathcal{E} G_{\text{ideal}},
\end{align}
which is an ideal quantum gate followed by some noise process $\mathcal{E}$. By inverting the ideal gate, we get the noise process as 
\begin{align}
\mathcal{E} = G_{\text{exp}}G_{\text{ideal}}^{-1}.
\end{align}
If we take the logarithm of this noise process and assume that noise is small \rev{i.e. $\mathcal{E}\simeq I$}. Using the approximation $\log X \simeq (X-I)$ with small $(X-I)$ we get the error generator \cite{blume-kohout_taxonomy_2022}
\begin{align}
L =\log(G_{\text{exp}}G_{\text{ideal}}^{-1})=\log\mathcal{E} \simeq \mathcal{E} - I,
\end{align}
which is the approximated difference between the noise process $\mathcal{E}$ and the identity. If the gate is noise-free, i.e., $\mathcal{E}=I$, then $L=0$. This error generator can be written into a linear combination
\begin{align}
L &= L_{H} + L_{S} + L_{C} + L_{A} \\
  &= \sum_{P} h_{P} H_{P} + \sum_{P} s_{P}S_{P} \\
  &+ \sum_{P,Q>P}c_{P,Q}C_{P,Q} + \sum_{P,Q>P} a_{P,Q}A_{P,Q}.
\end{align}
The terms in the linear combination correspond to error generators representing different error processes. The error generators are divided into four categories, Hamiltonian generator $H_{P}$, stochastic Pauli generator $S_{P}$, Pauli-correlation generator $C_{P, Q}$ and active generator $A_{P, Q}$. We use the $H+S$  model of pyGSTi package \cite{nielsen_probing_2020,madzik_precision_2022}, which only contains Hamiltonian and stochastic Pauli errors which have clear physical meanings. The Hamiltonian error generators represent a systematic over- or under-rotation of the qubit state on the Bloch-sphere in one of the rotation axes. On the other hand, the stochastic Pauli generators represent the contraction \rev{to one of the axes} of the qubit Bloch sphere. The coefficient of these error generator terms is obtained by \cite{blume-kohout_taxonomy_2022}
\begin{align}
h_{P} &= -\frac{i}{d^{2}} \Tr \left[ ( P \otimes I - I\otimes P) L \right], \\
s_{P} &= \frac{1}{d^{2}} \Tr \left[ (P-I) L \right] ,
\end{align}
where $P$ are the two-qubit Pauli matrices. We extract these coefficients using \lcc{internal} functions in the pyGSTi package. These coefficients can be used to calculate the Jamiolkowski probability $\epsilon_{J}(L)$ and the Jamiolkowski amplitude $\theta_{J}(L)$, which gives the amount of incoherent and coherent errors respectively. For an error generator $L$ being decomposed into list of $\{h_{P}, s_{P}\}$ coefficients, these two metrics are \cite{blume-kohout_taxonomy_2022}
\begin{align} 
\epsilon_{J}(L) &= \tr[\rho_{J}(L)(I-\ket{\Psi} \bra{\Psi})] = \sum_{P} s_{P}, \label{eq:J_prob}\\
\theta_{J}(L) &= \| (I-\ket{\Psi}\bra{\Psi})\rho_{J}(L)\ket{\Psi}\| = \left( \sum_{P} h_{P}^{2} \right)^{1/2}. \label{eq:J_amp}
\end{align}
The Jamiolkowski probability and Jamiolkowski amplitude can be used to approximate the averaged gate infidelity related with error generator $L$. For small errors, the approximated average gate infidelity is \cite{blume-kohout_taxonomy_2022}
\begin{equation}
r = \frac{d}{d+1}[\epsilon_{J}(L) + \theta_{J}(L)^{2}].
\end{equation}

\ifarXiv
    \foreach \x in {1,2, 3, 4, 5, 6}
    {
        \clearpage
        

\section*{Supplementary material}

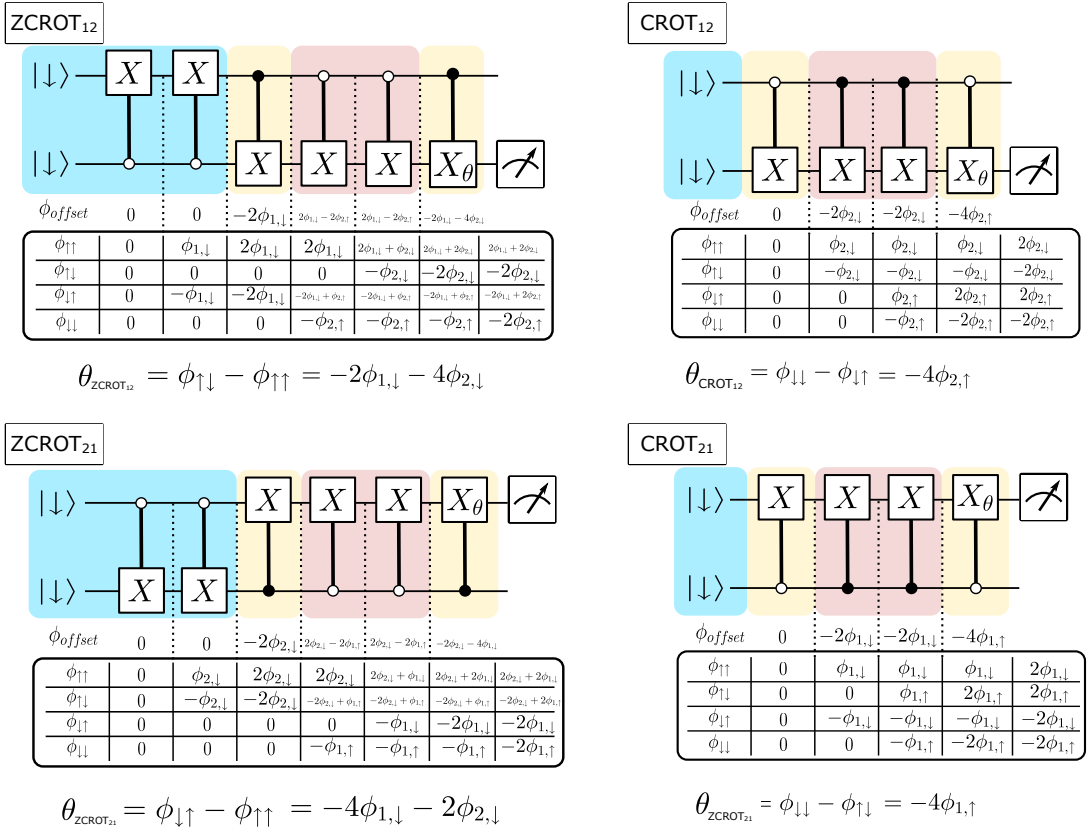

**Supplementary Figure S1: The four calibration sequences to measure the off-resonant Hamiltonian phase error.** All of the four calibration sequences used and tables which denote the phase errors accumulated along the sequences are shown. The sequences are composed of the preparation of the off-resonant state (blue), the Ramsey interference sequence (yellow) and the target gate (red). The phase error accumulated on each basis states before applying each pulse are shown in the columns. With the table of accumulated phases we obtain the phase offsets needed to compensate the effects of these phase errors. The relation between fitted shifts $\theta_i$ and the phase errors $\phi_{m,\sigma}$ can also be obtained with this table.

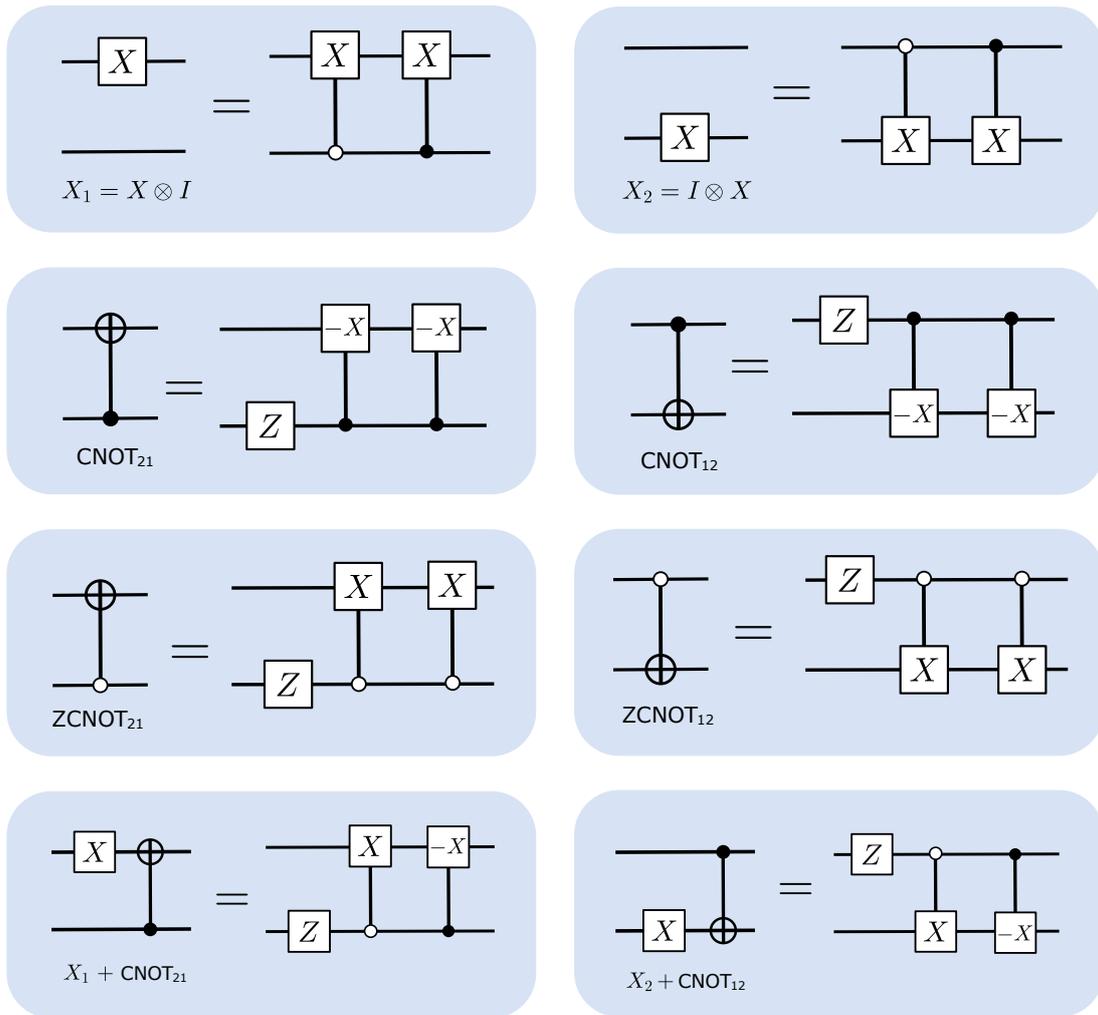

**Supplementary Figure S2: Decomposition of the two-Qubit randomized benchmarking primitive gates into CROT and ZCROT $\pi/2$ pulses**. The eight primitive gates used for the two-qubit randomized benchmarking experiment are decomposed with the four CROT and ZCROT pulses plus extra single qubit z-rotations [1, 2].

3

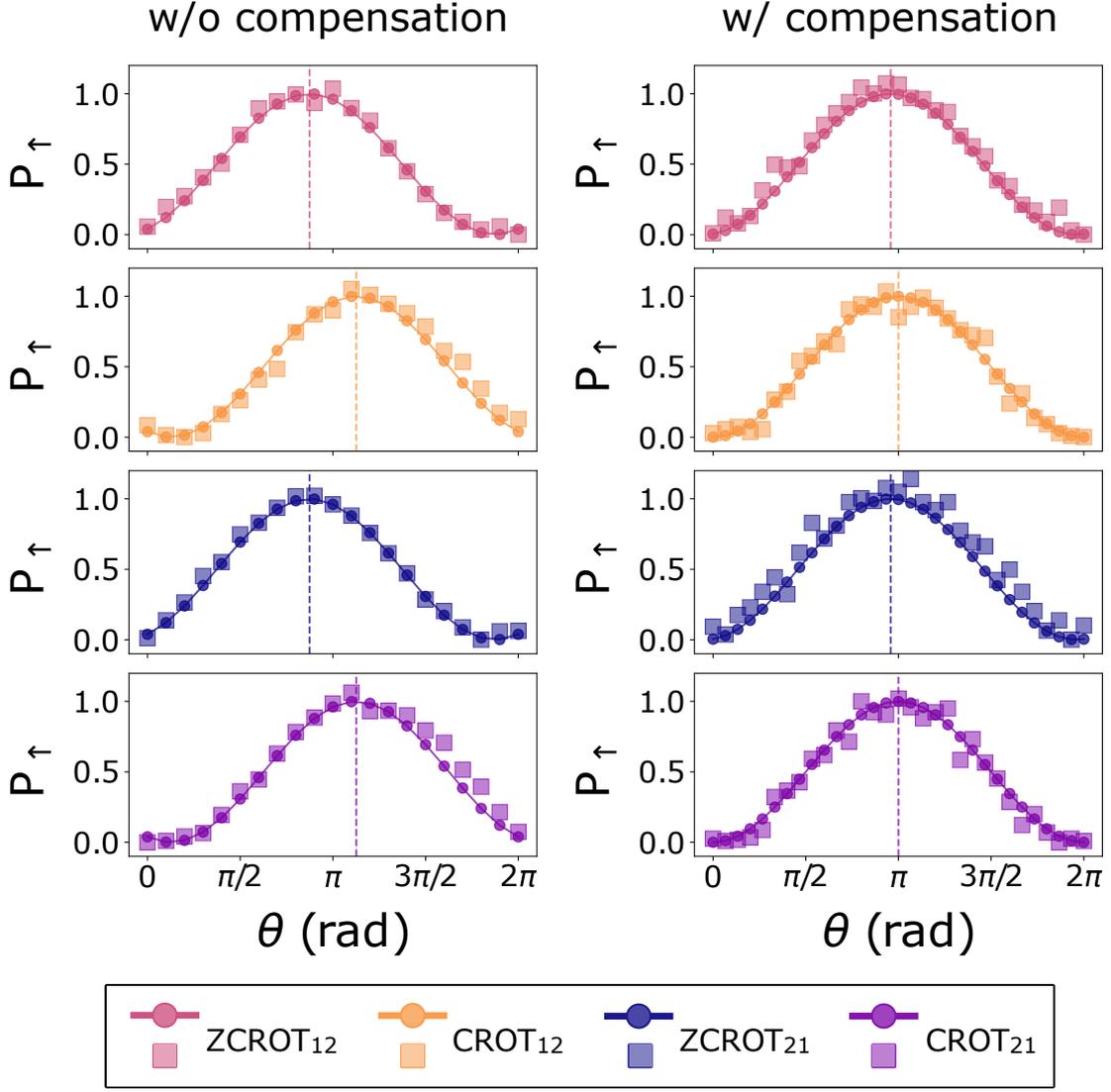

**Supplementary Figure S3: Simulated calibration sequences using ideal Hamiltonian with RWA**. The phase shifts fitted from the curve before compensation is $(\theta_{\mathrm{ZCROT}_{12}}, \theta_{\mathrm{CROT}_{12}}, \theta_{\mathrm{ZCROT}_{21}}, \theta_{\mathrm{CROT}_{21}}) = (0.396, -0.396, 0.396, -0.396)$ rad. The four off-resonant Hamiltonian phase errors are then $(\phi_{1,\downarrow}, \phi_{1,\uparrow}, \phi_{2,\downarrow}, \phi_{2,\uparrow}) = (-0.066, 0.1, -0.066, 0.1)$ rad. After compensation the shifts becomes $(\theta'_{\mathrm{ZCROT}_{12}}, \theta'_{\mathrm{CROT}_{12}}, \theta'_{\mathrm{ZCROT}_{21}}, \theta'_{\mathrm{CROT}_{21}}) = (0.132, 0.000, 0.132, 0.000)$ rad. All the errorbars of the fitted phases are smaller then $10^{-6}$ rad. The two ZCROT sequences shows residual phase shifts after compensation as discussed in the main text. The normalized probability obtained from the experiment is shown as squares in the plots. By using experimentally obtained parameters, our simulation fits the experimental data well.

4

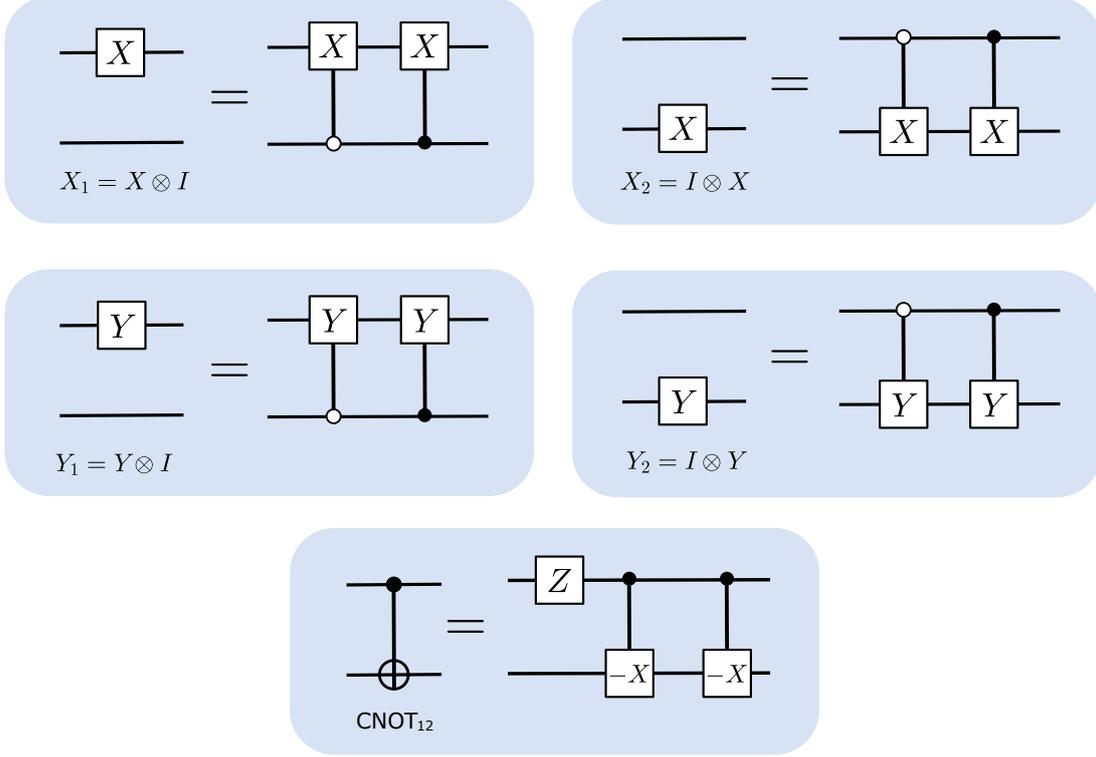

**Supplementary Figure S4: Composition of quantum gates with CROT and ZCROT $\pi/2$ pulses**. The quantum gates used in GST are all composed of two half-$\pi$ pulses. The Y gates are implemented with the same parameters as the X gates but shift the microwave phase by $\pi/2$. The virtual z half-$\pi$ rotation is executed to obtain the logical CNOT$_{12}$ gate.

| Gate | Avg. Gate Infidelity | Non-unitary Avg. Gate Infidelity | Jamiolkowski Probability | Jamiolkowski Amplitude | 1/2 Diamond-Dist |
|---|---|---|---|---|---|
| $I$ | 0.0093 ± 0.0037 | 0.0093 ± 0.0036 | 0.0117 | 0.0083 | 0.0173 ± 0.0076 |
| $X_1$ | 0.0050 ± 0.0016 | 0.0041 ± 0.0017 | 0.0052 | 0.0335 | 0.0473 ± 0.0011 |
| $Y_1$ | 0.0081 ± 0.0012 | 0.0075 ± 0.0013 | 0.0095 | 0.0270 | 0.0390 ± 0.0151 |
| $X_2$ | 0.0106 ± 0.0019 | 0.0102 ± 0.0019 | 0.0129 | 0.0230 | 0.0358 ± 0.0106 |
| $Y_2$ | 0.0055 ± 0.0011 | 0.0051 ± 0.0011 | 0.0064 | 0.0226 | 0.0333 ± 0.0087 |
| CNOT$_{12}$ | 0.0052 ± 0.0067 | 0.0045 ± 0.0068 | 0.0056 | 0.0292 | 0.0395 ± 0.0062 |

**Supplementary Table S1: GST gauge dependent metrics of the gate-set $\{I, X_1, Y_1, X_2, Y_2, CNOT_{12}\}$ with the phase error compensation protocol.** The averaged gate infidelity $\frac{\text{tr}(G_{\text{exp}}^{-1} G_{\text{ideal}}) + d}{d(d+1)}$, non-unitary averaged gate infidelity $\frac{d-1}{d}\left(1 - \sqrt{\frac{\text{tr}(G_{\text{exp, u}}^{\dagger} G_{\text{exp,u}})}{d^2 - 1}}\right)$, Jamiolkowski probability [Eq. (30)], Jamiolkowski amplitude [Eq. (31)] and the one-half diamond distance $\frac{1}{2}\max_\rho \|(G_{\text{ideal}} \otimes I)\rho - (G_{\text{exp}} \otimes I)\rho\|_1$ are shown.

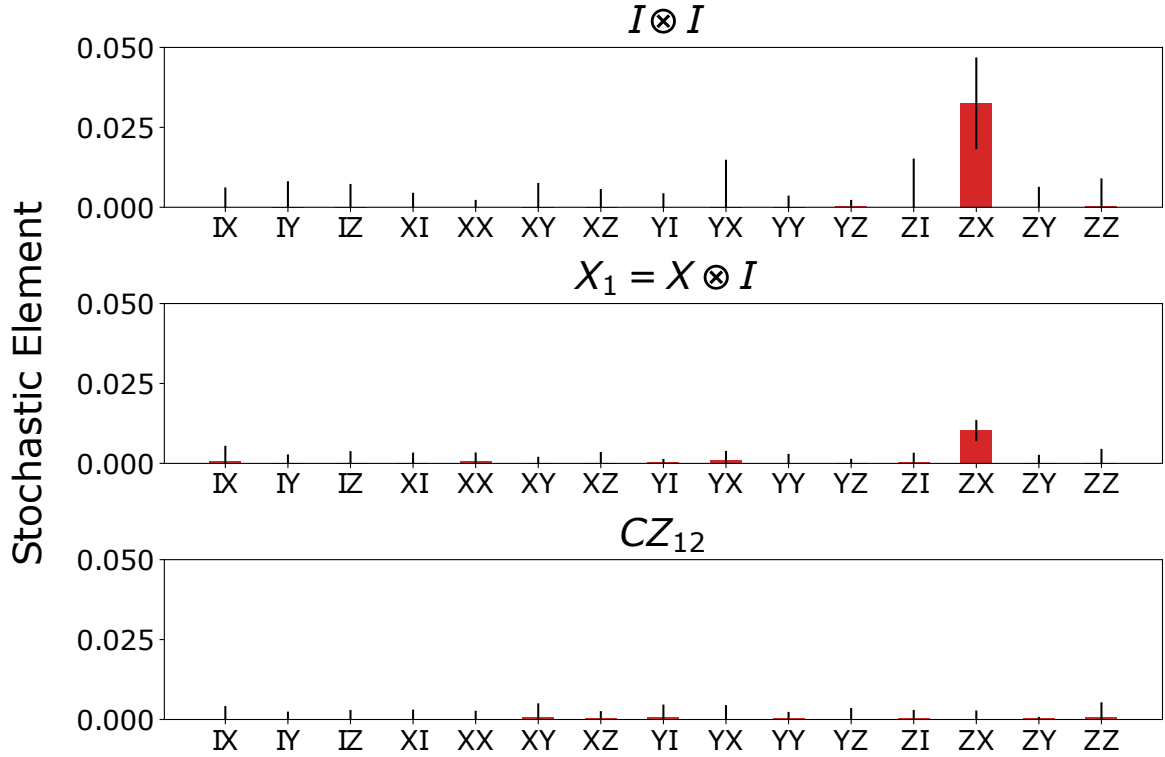

**Supplementary Figure S5: Stochastic noise component of the virtual CZ gate-set.** Full data set of data shown in Fig. 4 (c) of main article. Error bars represent $1\sigma$ standard deviation from the mean.

# Supplementary References

[1] Noiri, A. *et al.* Fast universal quantum gate above the fault-tolerance threshold in silicon. *Nature* **601** (7893), 338–342 (2022) .

[2] Huang, W. *et al.* Fidelity benchmarks for two-qubit gates in silicon. *Nature* **569** (7757), 532–536 (2019) .


    }
\fi
\end{document}